\documentclass[useAMS,usenatbib]{mnras}
\PassOptionsToPackage{hyphens}{url}\usepackage{hyperref}
\usepackage{epsfig,lscape} 
\usepackage{hyperref} 
\usepackage{natbib}
\usepackage[caption=false]{subfig}
\usepackage{lscape}
\usepackage{textcomp}
\usepackage{gensymb}
\usepackage{xcolor}
\usepackage{amsmath}
\usepackage{comment}
\usepackage{arydshln}
\usepackage{graphicx}
\usepackage{booktabs}
\usepackage[caption=false]{subfig}
\usepackage{amsmath,amssymb}
\usepackage[normalem]{ulem}

\title[Neutron-capture element evolution]{Observational constraints on the origin of the elements. VI. Origin and evolution of neutron-capture elements as probed by the Gaia-ESO survey}

\author[J. Lian et al.]{
{Jianhui Lian}$,^{1,2}$\thanks{jianhui.lian@ynu.edu.cn} 
{Nicholas Storm},$^{1}$ 
{Guillaume Guiglion},$^{1}$ 
{Aldo Serenelli},$^{3,4}$  
{Benoit Cote},$^{6,7}$ 
\newauthor {Amanda I. Karakas},$^{8,9}$
{Nick Boardman},$^{10}$ 
and {Maria Bergemann}$^{1,5}$
\\
$^{1}$Max Planck Institute for Astronomy, 69117, Heidelberg, Germany \\
$^{2}$ South-Western Institute for Astronomy Research, Yunnan University, Kunming, Yunnan 650091, People’s Republic of China \\
$^{3}$Institute of Space Sciences (ICE, CSIC), 08193, Cerdanyola del Valles, Spain \\
$^{4}$Institut d'Estudis Espacials de Catalunya (IEEC), 08034, Barcelona, Spain \\
$^{5}$ Niels Bohr International Academy, Niels Bohr Institute, Copenhagen, Denmark \\
$^{6}$ Department of Physics and Astronomy, University of Victoria, Victoria, BC V8P 5C2, Canada  \\
$^{7}$ Joint Institute for Nuclear Astrophysics—Center for the Evolution of the Elements (JINA-CEE), USA\\ 
$^{8}$ School of Physics \& Astronomy, Monash University, Clayton VIC 3800, Australia \\
$^{9}$ 2ARC Centre of Excellence for All Sky Astrophysics in 3 Dimensions (ASTRO 3D)\\
$^{10}$ School of Physics and Astronomy, University of St Andrews, North Haugh, St Andrews KY16 9SS, UK \\
}

\begin{document}

\maketitle

\begin{abstract}
{Most heavy elements beyond the iron peak are synthesized via neutron capture processes}. The nature of the {astrophysical} sites of neutron capture processes {is} still very unclear. 
In this work we explore the observational constraints of the chemical abundances of s-process and r-process elements on the sites of neutron-capture processes by applying {Galactic} chemical evolution (GCE) models to the data from Gaia-ESO {large spectroscopic stellar} survey. For the r-process, the [Eu/Fe]-[Fe/H] distribution suggests a short delay time of the site that produces Eu. Other independent observations (e.g., NS-NS binaries), however, suggest a significant fraction of long delayed ($>1$~Gyr) neutron star mergers (NSM). When assuming NSM as the only r-process sites, these two observational constraints are inconsistent at above 1$\sigma$ level. Including short delayed r-process sites like {magneto-rotational} supernova can resolve this inconsistency. For the s-process, we find a weak metallicity dependence of {the} [Ba/Y] ratio,  which traces the s-process efficiency. Our GCE model with up-to-date yields of AGB stars qualitatively reproduces this metallicity dependence, {but the model predicts a much higher [Ba/Y] ratio compared to the data.} This mismatch suggests that the s-process efficiency of low mass AGB stars in the current AGB nucleosynthesis {models could be} overestimated.  

\end{abstract}


\section{Introduction}

Studies of chemical structure of the Milky Way have seen a long history of research. However, most progress so far has been made towards understanding the cosmic production of low-Z (Z $<$ 30) elements, which accessible in spectral windows targeted by ongoing large spectroscopic surveys, such as APOGEE, LAMOST, GALAH. In contrast, little is known about the detailed Galactic chemical evolution {(hereafter, GCE)} and formation sites of elements beyond Fe (Z $>$ 30), which are formed by captures of neutrons on seed nuclei (e.g., \citealt{reifarth2014}, \citealt{cowan2021}). Many of the isotopes are formed by the s-process, involving slow capture of neutrons and yielding nuclei close to the valley of stability. The s-process is typically attributed to asymptotic giant branch (AGB) and massive stars (e.g., \citealt{arlandini1999,herwig2005,straniero2006,karakas2014}). The weak s-component, which comprises nuclei with Z $<$ 39, can be synthesised in He-core and C-shell burning in massive stars (e.g., \citealt{pignatari2010,frischknecht2016,choplin2018}). The main s-process, which produces species 39 $<$ Z$<$ 81, is thought to occur in $^{13}$C-rich pockets in AGB stars (e.g., \citealt{karakas2014,cristallo2015}). Also, the LEPP (Light Element Primary Process) has been proposed to operate at low metallicity \citep{travaglio2004,montes2007}. 

Other elements are made via nuclear transmutations following rapid capture of neutrons, in the so-called r-process. The r-process synthesis involves processes operating far from the nuclear stability valley and, thus, requires extreme conditions. The sites include neutrino-driven high-entropy winds in core collapse SNe \citep[e.g.,][]{farouqi2010,arcones2013,bliss2018a}, explosions of rapidly-rotating magnetised massive stars \citep[e.g.,][]{siegel2017,halevi2018,siegel2019,reichert2023}, compact binary mergers of two neutron stars (NS; e.g., \citealt{freiburghaus1999,korobkin2012,martin2015,radice2018}), and a merger of a neutron star and a black hole (BH; e.g., \citealt{lattimer1974,wehmeyer2019}). Also {quark deconfinement SNe II \citep{fischer2020} have been proposed as r-process sites, although their existence is still debated \citep{cowan2021}.}
Some of the heavy nuclei are only produced by charged-particle reactions involving protons \citep[e.g.,][]{arnould2003,nishimura2018,travaglio2018,battino2020}. These conditions are also found in core-collapse SNe and in SN Ia. Ultimately, some elements could be produced in super-AGB stars \citep{jones2016} or in rapidly accreting white dwarfs \citep{cote2018a}. Up to 45$\%$ of Rb, Sr, Y, Zr, and Mo could be produced in {the latter scenario \citep{cote2018a}}. \citet{bliss2018b}, on the other hand, find that Mo and Ru could be explained by proton-rich neutrino-driven winds from nascent neutron stars. Whereas one process or the other can preferentially make some isotopes, most nuclei are made by both s- and r-processes, thus detailed studies of the potential sites of neutron-capture elements must self-consistently include all these sources in GCE models. 

In this work, we aim to explore the potential of observed abundances in Galactic populations to constrain the formation and origins of neutron-capture elements and their production cites. We make use of the chemical abundances in {the disc and halo} stars from the final {public} data release of the Gaia-ESO large spectroscopic survey \citep{gilmore2022,randich2022} {and analyse the Galactic distributions in the context of} predictions of the GCE model OMEGA+ \citep{cote2018} that allows for very flexible treatment of sites and yields. 
{We} combine observational information for seven elements: six s-process elements (Y, Zr, Ba, La, Ce, Nd) and one r-process element (Eu), {for which detailed} abundances are available in Gaia-ESO survey \citep{gilmore2022,randich2022}. {We compare the observed distributions with the GCE model predictions and attempt to assess the potential of such combined studies in the context of large-scale spectroscopic surveys.}

\section{Observations}
\begin{figure}
\centering
\includegraphics[width=0.48\textwidth]{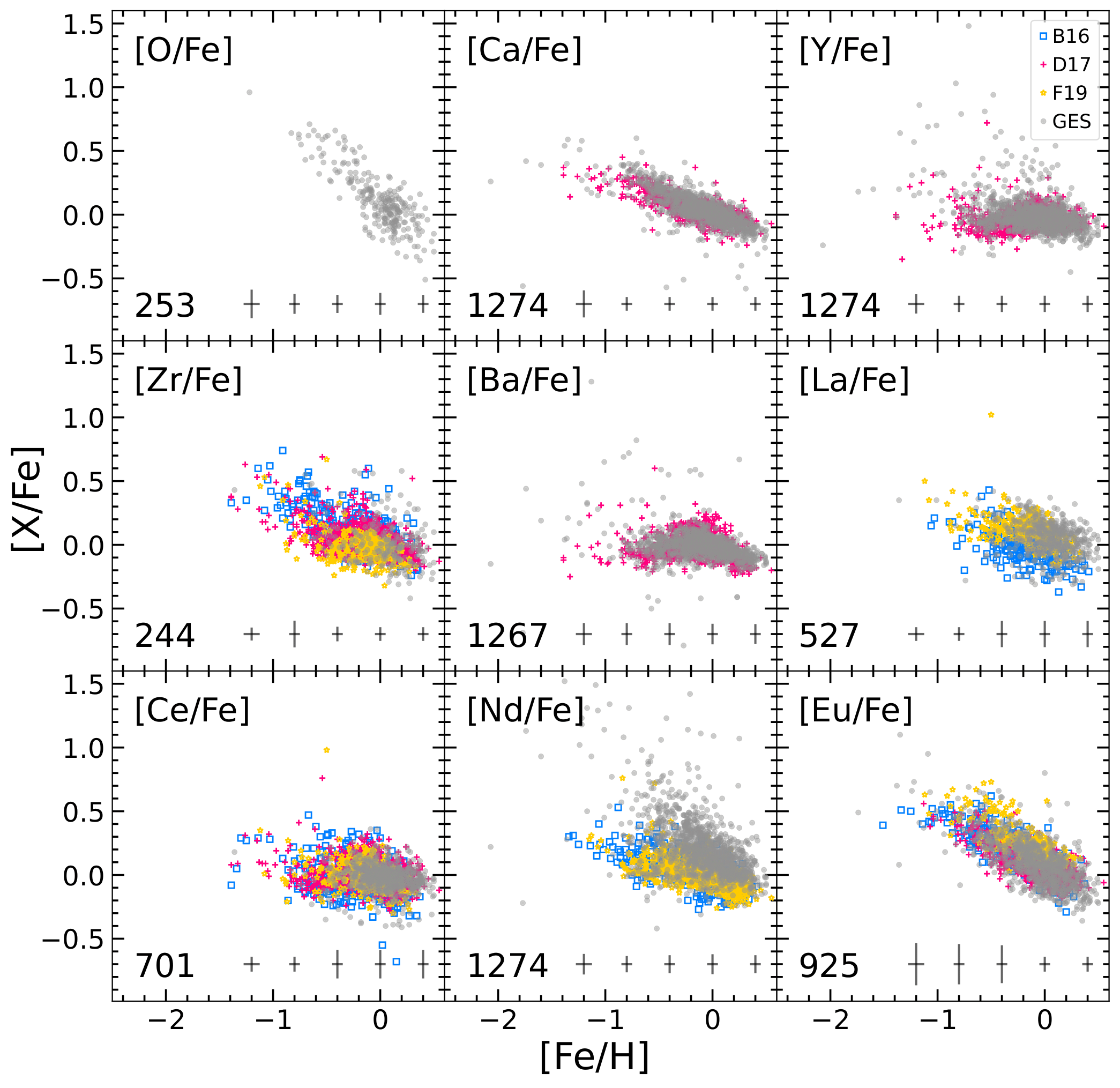}
\caption{[X/Fe]-[Fe/H] diagram for seven neutron-capture elements with reliable detection in Gaia-ESO survey. These seven elements are grouped into three categories: light s-process elements (Y, Zr) in the left two panels at the top, heavy s-process elements (Ba, La, Ce, Nd) at the bottom, and r-process element (Eu) in the right panel at the top. For comparison, three other samples with abundance measurement of these neutron-capture elements from previous high-resolution spectroscopic works \citep{battistini2016,delgado2017,forsberg2019} are included.} 
\label{xfe-feh-ges}
\end{figure} 
In this work we use data from the latest {public data} release of the Gaia-ESO survey \citep{gilmore2022,randich2022}. {Gaia-ESO was an optical, medium- and high-resolution spectroscopic survey at the 8-meter Very Large Telescope (VLT) of European Southern Observatory (ESO) that acquired high-quality spectra for hundreds of thousands of stars in the Galactic disc, bulge and halo, as well as in open and globular clusters}. The relatively wide wavelength coverage and high spectral resolution enable robust measurement of stellar parameters and dozens of elemental abundances, including $\alpha$, odd-Z, iron-peak, s- and r-process neutron capture elements \citep{magrini2018}. The element abundances are estimated {in 1D LTE (local thermodynamic equilibrium) and 1D NLTE (non-local thermodynamic equilibrium), using the MARCS model atmospheres \citep{gustafsson2008} and the dedicated Gaia-ESO line list \citep{heiter2021}. Different spectral analysis methods are used to minimize the statistical error (for a more detailed description, see \citealt{smiljanic2014} and \citealt{sacco2014})}.
The abundances are given in absolute units, i.e. A(X) = 12+log10(N$_{\rm X}$/N$_{\rm H}$). To convert into conventionally solar-normalised values, we adopt the solar abundance pattern from \citet{grevesse2007}.

The selection of our sample is limited to {the Galactic field stars. For a detailed description of the observing strategy and selection of fields we refer the reader to \citet{Stonkute2016}}. {We further limit the sample to stars with {modest} uncertainties in the measurements of stellar parameters and neutron-capture elements to ensure reliable estimates of stellar parameters and elemental abundances. Our final Gaia-ESO sample thus contains 1274 stars satisfying the following criteria}:
\begin{itemize}
    \item GES\_TYPE set to GE\_MW;
    \item SNR $>$ 50;
    \item $\sigma_{\rm teff}<100$~K, $\sigma_{\rm logg}<0.2$;
    \item $\sigma_{\rm Y}<0.2$, $\sigma_{\rm Nd}<0.2$. 
\end{itemize}
Figure~\ref{xfe-feh-ges} shows the distribution of {our Gaia-ESO sample in [X/Fe]-[Fe/H] plane for two $\alpha$ elements (O and Ca) and seven neutron-capture elements detected in the Gaia-ESO spectra. This includes six s-process {dominated} elements (Y and Zr in the first peak and Ba, La, Ce, and Nd in the second peak) and one r-process {dominated} element (Eu). We emphasize that none of these elements can be characterised as strictly s- or r-process, as different isotopes of the same element are produced in different nucleosynthetic processes \citep[e.g.][]{prantzos2020}. So, for example, Nd is a mixed element, with a contribution of main s-process at the solar age of about 57 \% only \citep{travaglio2004,Bisterzo2014}.}
For comparison, we also include other measurement of neutron-capture elements of different {Galactic} samples compiled from previous works \citep{battistini2016,delgado2017,forsberg2019}. For s-process elements, their abundance with respect to Fe is overall around the solar value. Interestingly, there is a weak bump feature in {Y and Ba with a local maximum at a slightly sub-solar metallicity ([Fe/H]$\sim$-0.2). Eu - an element dominated by r-process nucleosynthesis} - behaves {similarly to oxygen}, showing a rapid monotonic decrease from [Eu/Fe]$\sim+0.5$ 
at [Fe/H]$\sim$-1 to [Eu/Fe]$\sim0$ at [Fe/H]$\sim$0.2. These trends are also seen in the samples of previous studies, suggesting {that} {these elements have robust observational detections, although it should be noted that all these s- and r-process patterns refer to 1D LTE measurements and systematic uncertainties incurred by these assumptions remain to be quantified by detailed self-consistent calculations with NLTE and 3D models \citep[e.g.][]{bergemann2014_nlte,bergemann2019}.} 
\section{Chemical evolution model}
{Galactic} chemical evolution (GCE) modeling is a useful tool to interpret the observed elemental abundance pattern of stellar populations to understand the production and evolution of heavy elements as well as the implied galaxy formation and evolution history. 
\subsection{Basic ingredients}\label{sec:gce}
Since the goal of this work is to explore the constraining power of spectroscopic stellar surveys on the {production of heavy} elements, we use the OMEGA+\footnote{https://github.com/becot85/JINAPyCEE} {GCE}  \citep{cote2017,cote2018} for its high flexibility in treating chemical enrichment sources. This model considers star formation, gas inflow and outflows. For an in-depth discussion of these parameters and their role in GCE, we refer the reader to excellent literature studies {\citep[e.g.][and references therein]{Matteucci2021}}. Here, we will only summarizes the choices relevant to our model.

The star formation is assumed to follow the {Kennicutt-Schmidt} (KS) star formation law \citep{1959ApJ...129..243S,kennicutt1998}, in the commonly used form \citep[e.g.][eq. 4]{cote2017} $\dot{M_*}(t)=\epsilon_{*}\frac{M_{\rm gas}(t)}{\tau_{*}}$, where $\dot{M_*}$ is the star formation rate (SFR), $\epsilon_{*}$ is the star formation efficiency, and $\tau_*$ is the star formation timescale in years, i.e. how long it takes for neutral gas to be converted into stars. 
Therefore the ratio between the SFR and gas mass is regulated by both $\epsilon_{*}$ and $\tau_*$ in the model. We adopt the default configuration of the star formation model in OMEGA+, i.e., a constant $\epsilon_{*}$ and an evolving $\tau_*$, which is assumed to be proportional to the dynamical timescale ($\tau_{\rm dyn}$) of the virialized system including both dark matter and baryons (see \citealt{cote2017}, their Eq.~(12) and references {therein}). We assume a simplified form of the gas accretion history, with the gas rate declining exponentially with an e-folding time $\dot{M}_{\rm{in}}(t) = a e^{-t/{\tau_{\rm inflow}}}$, where $a$ is a normalisation constant and the inflow timescale for a basic model is set to $\tau_{\rm inflow} = 7$~Gyr. For the gas outflow {rate}, which is modelled as $\dot{M}_{\rm{out}}(t) = \eta \dot{M_*}(t)$, we adopt a constant mass loading factor $\eta$ of 1. {That is, the gas outflow rate is equal to the star formation rate. The initial mass function (IMF) is adopted from \citet{kroupa2001}}. A detailed discussion underlying the choice of the free parameters in the model can be found in \citet{cote2017} and \citet{cote2018}. We note that this star formation history (SFH) may not be representative of the detailed multi-phase evolutionary history of the Galactic disc \citep[e.g.,][]{schonrich2019, spitoni2019, lian2020}. However, as we will demonstrate in {Sect. \ref{sec:res}}, the main observed features in the {evolution of abundance patterns of} of neutron-capture elements are mostly due to the nature of their formation sites {and the choice of SFH parameters in the model only weakly influences the overall trends of the chemical abundance ratios}. Thus adopting a more sophisticated star formation history will not change the results of this work qualitatively. 

\subsection{Yields}\label{sec:yields}
A variety of tables of yields from different sites of heavy elements is available in OMEGA+. Here we summarize the yields table of each site that we adopt in the GCE model. 

{\bf AGB stars}: An important late evolution stage of intermediate and low mass ($\sim$1-8${\rm M_{\odot}}$) stars is the thermally-pulsing AGB phase. During this phase, the alternate burning of the H-shell and the unstable He-shell leads to the occurrence of thermal pulses, followed by third-dredge up events which drive the mixing between the shells where nuclear reactions take place and the surface, enriching the latter with carbon and s-process elements \citep{busso1999}. AGB stars experience intense mass loss (from $10^{-7}$ to $10^{-4}{\rm M}_{\odot}{\rm yr^{-1}}$, \citealp{habing1996,groenewegen:2018}) that carry enriched material back into the surrounding interstellar medium. There are multiple yields tables of AGB stars available in {OMEGA+}. In this work we adopt the table from \citet{cristallo2015} in our basic model and another table from \citet{karakas2010} to test the impact of choice of AGB yields tables on the enrichment of s-process elements. The latter only contains yields up to Ni. Recently, the authors derived updated AGB yields that include neutron-capture elements based on new calculations of AGB evolution and nucleosynthesis at different metallicities from $-2$ to $0.3$ \citep{karakas2014,karakas2018,karakas2022}. We adopt the updated Karakas's AGB yields in the GCE model. For AGB stars with masses less than 4~${\rm M_{\odot}}$, we adopt the yields tables that assume partial mixing zone  (PMZ) with mass of 10$^{-3}\ \rm M_{\odot}$; for AGB stars with masses between 4 and 5~${\rm M_{\odot}}$, we adopt the yields that assume PMZ mass of 10$^{-4}\ \rm M_{\odot}$; and for AGB stars above 5~${\rm M_{\odot}}$, no PMZ is assumed. 
%
%

{\bf Type Ia (SN Ia) supernova}: OMEGA+ includes several SN Ia channels. One of them is the canonical Chandrasekhar-mass (M$_{\rm ch}$) channel representing a single-degenerate binary system (one white dwarf accreting H- or He-rich material from a non-degenerate companion) that end up in thermonuclear explosion, which is modelled using the yields table from \citet{iwamoto1999}. There is a growing evidence that sub-(M$_{\rm ch}$) explosions of single- or double-degenerate white dwarfs also make a significant contribution to the chemical evolution of Fe-peak elements, such as Mn \citep{seitenzahl2013,kobayashi2020,eitner2020,sanders2021,eitner2022}. Following \citet{eitner2022}, we consider four main types of SN Ia scenarios, two with (M$_{\rm ch}$) explosions and two with sub-(M$_{\rm ch}$) explosions, with 62\% of SN-Ia {(in terms of the total number of SN Ia events)} stem from sub-(M$_{\rm ch}$) channel. {The SN Ia types considered include the single-degenerate near-Chandrasekhar mass SN-Ia, the fainter SN-Iax systems associated with He accretion from the companion, as well as two sub-Chandrasekhar mass channels associated with the double-detonation of a white dwarf accreting helium-rich matter and violent mergers of white dwarf binaries.} The yield tables of each type of SN-Ia are taken from the Heidelberg Supernova Model Archive \citep{kromer2017}. The delay time distributions {(DTD)}, which define the number of events since the formation of the progenitor systems, are different for each of these SN-Ia scenarios (see Fig.~3 in \citealt{eitner2022}). {They are modelled using the \textsc{StarTrack} {binary} population synthesis code.}

{\bf Core collapse supernovae (CCSN) and stellar winds:} Core collapse supernova, the {final stage of the evolution of} massive stars (roughly 10-25 ${\rm M_{\odot}}$), is one of the major sources of {Fe-group elements}. However, the contribution of CCSN to the production of neutron-capture elements is still under debate. Rotation of the progenitor massive stars {has been} shown to be an important parameter for the production of s-process elements \citep{limongi2018}. CCSN resulting from the evolution of {rapidly rotating massive stars (rotation velocity of 300~km/s on the main-sequence) produce abundances of s-process elements that are orders of magnitude higher than systems with slower rotation velocity of 150 and 0~km/s, particularly at sub-solar metallicities. This is a result of rotationally induced mixing that transports $^{14}$N produced in the H-burning shell down to the He-core where it builds up the $^{22}$Ne content, the main neutron source for s-process in this environment.} \citet{prantzos2018} obtained an empirical rotation velocity distribution of CCSN (depending on metallicity) to avoid over-production of s-process elements {compared to observations}. {In this work we adopt the recommended yields from \citet[][their set "R"]{limongi2018}. These dataset represent yields from stars with the initial mass 13--120 ${\rm M_{\odot}}$ at four values of metallicity [Z]$=-3, -2, -1, 0$ and three rotation velocities of $0, 150, 300$~km/s. For stars with the initial mass above 25 ${\rm M_{\odot}}$, only enrichment due to stellar winds is relevant, as these systems are assumed to end as black holes.} We obtain the average yields of massive stars in terms of rotation using the rotation velocity distribution derived by \citet{prantzos2018}. 

\textbf{Magneto-rotational supernova} (MRSN): In addition to fast rotation, massive stars with strong magnetic fields have also been suggested {as sites of neutron-capture elements} (in particular r-process) compared to normal CCSN \citep[e.g.,][]{cameron2003,nishimura2015,yong2021,reichert2023}. 
This rare type of {explosions} has been incorporated in GCE models in several studies to explain the observed Eu abundance in disk and halo stars \citep[e.g.,][]{cescutti2014,cote2019,simonetti2019,kobayashi2020}. However, MRSN is unlikely to be the only site of r-process elements, as {a possible signature of r-process elements is detected in the NS-NS merger event GW170817 }\citep[e.g.][]{watson2019,perego2022}. In this work, we incorporate MRSN in OMEGA+ following {an approach similar to that of} \citet{kobayashi2020}, by replacing a very small fraction of normal CCSN (progenitor mass range in 13-25~${\rm M_{\odot}}$) by MRSN. We find a fraction of 0.25\% provides a good match to the observed [Eu/Fe] as will be shown in \textsection4.3. 
The yields of MRSN are adopted from the work of \citet{nishimura2015}, where the nucleosynthesis yields is calculated as a post-processing based on magneto-hydrodynamic simulations of a 25 ${\rm M_{\odot}}$ star \citep{takiwaki2009}.

{\bf Neutron star mergers (NSM):} {The merger origin of r-process elements is supported by the recent observations of the kilonova associated with the gravitational wave detection GW170817 \citep{Abbott2017a}. 
A significant amount of {lanthanide} elements are needed to explain the observed kilonova spectrum and light curve \citep[e.g.,][]{chornock2017,cowperthwaite2017,drout2017,pian2017,tanaka2017,villar2017}. 
According to simulations , the ejected mass of r-process elements by NSM is of the order} of $10^{-2}~{\rm M_{\odot}}$ (see Table 1 in \citealt{cote2018} and references therein), which depends on the total mass and mass ratio of the merging neutron stars \citep{korobkin2012,bovard2017}. In this work, we follow the default configuration of NSM in OMEGA+ that assumes an ejecta mass of NSM of 2.5$\times 10^{-2} {\rm M_{\odot}}$, power-law DTD, and yields table from \citet{arnould2007}.  
The rate of NS-NS mergers in the local Universe is estimated to be $1540^{+3200}_{-1220}{\rm Gpc^{-3}yr^{-1}}$ \citep{abbott2017c}. This occurrence rate is suggested to be high enough to explain the entire r-process mass in the Milky Way \citep{abbott2017c}. {However, it shall be noted that the event GW170817 is detected in an early-type galaxy \citep{abbott2017d} with little on-going star formation \citep{blanchard2017,levan2017,pan2017}. It is possible that the delay time of this merger event is greater than a few Gyr \citep{blanchard2017,pan2017}.}

{We note that the NSM yields from \citet{arnould2007} {are} obtained based on the r-process residual distributions derived by \citet{goriely1999}, which may not be fully consistent with the more recent calculations of s-process yields of AGB stars and CCSN. {Whereas this choice might be relevant when considering the relative abundance pattern of r-process isotopes, it is not critical within the scope of the GCE analysis in this work.} Specifically, for Eu isotopes, the r-process residuals from \citet{goriely1999} are in good agreement with those derived by \citet{prantzos2020}, which are fully consistent with AGB yields from \citet{cristallo2015} and CCSN yields from \citet{limongi2018}. Also, concerning the enrichment history of [Eu/Fe], the release timescale of Eu from different sites is more important than the amount of Eu released by each event, given the large uncertainties in the occurrence rates of these events. For these reasons, we keep using the NSM yields from \citet{arnould2007} in our GCE model.} 
%
\begin{figure}
\includegraphics[width=0.46\textwidth]{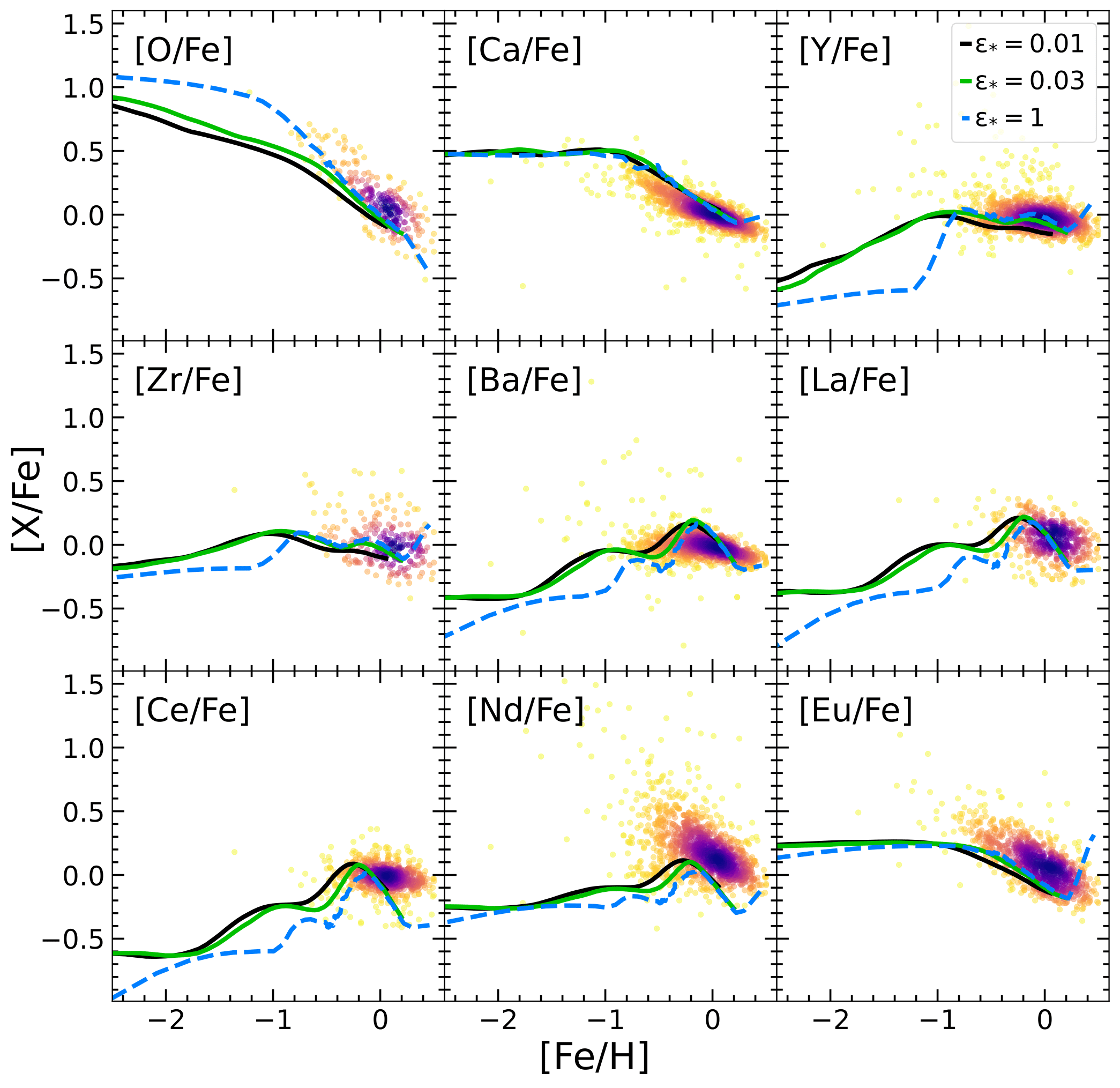}
\caption{Comparison of GCE tracks calculated with different star formation efficiency parameters $\epsilon_*$ with observations.}
\label{sfh_sfe}
\end{figure} 

\begin{figure}
\includegraphics[width=0.46\textwidth]{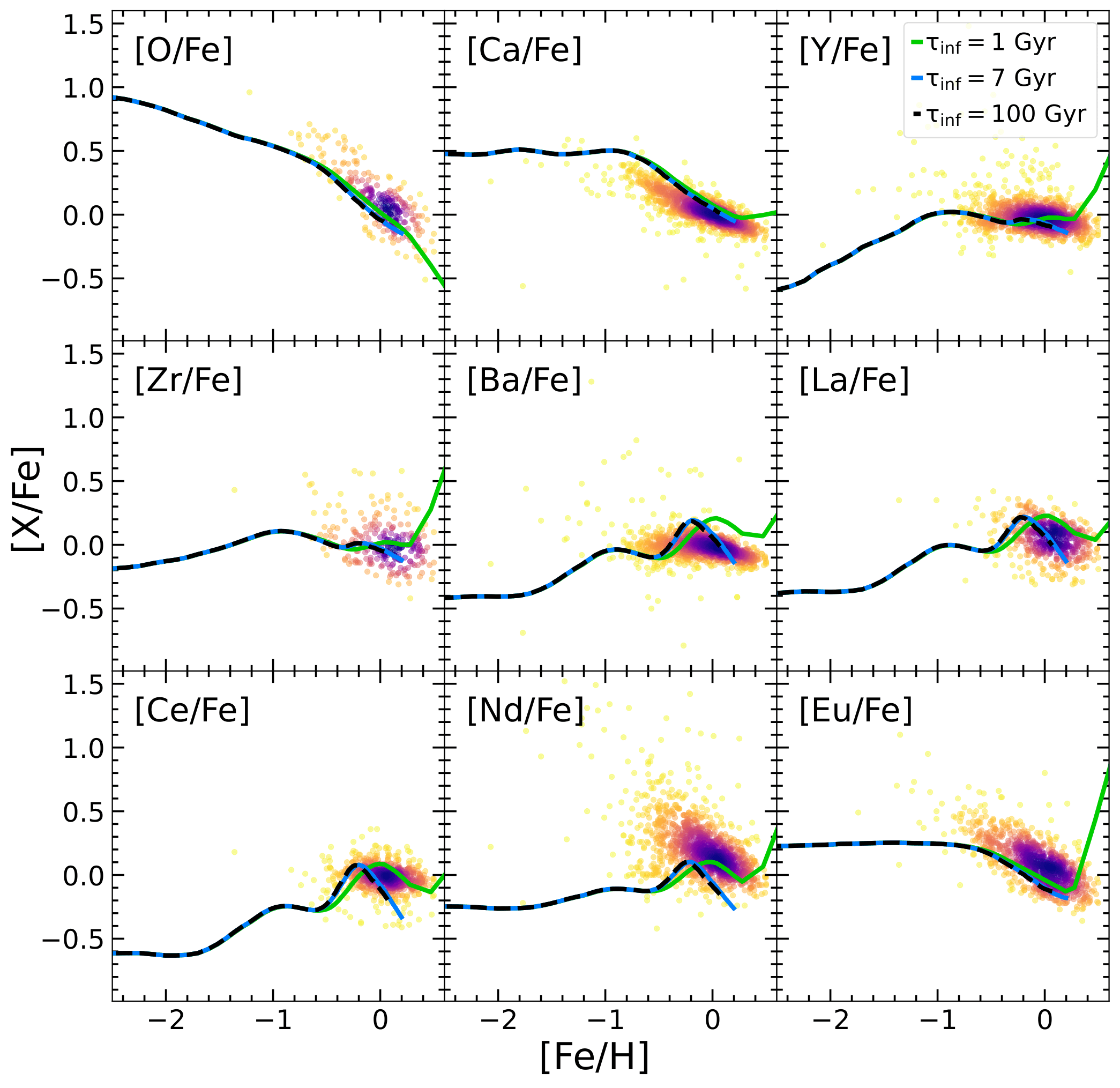}
\caption{Comparison of GCE tracks calculated with different inflow timescales $\tau_{\rm inflow}$ with observations.}
\label{sfh_inflow}
\end{figure} 
\section{Results}\label{sec:res}
In this section we confront the GCE model with the observed abundances of s- and r-process elements to extract observational constraints on their production sites and origin. We explore the parameter space of the GCE model, including SFH, yields tables, and sites. We begin with the analysis of the influence of the star formation history and gas accretion timescale, and then proceed with the studies of individual enrichment channels.
%
%
%
%
\subsection{Influence of SFH}  
The SFH therefore plays a fundamental role in the galactic enrichment history of metals, including the heavy neutron-capture elements. Since we adopt a simplified SFH framework in this work, here we first explore the influence of {the adopted SFH parameterisation (Sect. \ref{sec:gce})}. Here we assume our basic model that includes the contribution of CCSN, SN-Ia, AGB, MRSN, and NSM in the chemical evolution.

Figure \ref{sfh_sfe} shows {the chemical GCE tracks calculated with different SFE ($\epsilon_{\rm *}$)} in comparison with the observed abundance distributions. {The lighter elements, O and Ca are affected in a slightly different way. Ca is invariant to the modification of the star formation efficiency, whereas O is produced slightly more efficiently for higher values of $\epsilon*$. But the differences in the corresponding GCE curves are only clearly detectable at [Fe/H] $\lesssim -1$, and hence are barely constrained by available observational data.} {For the s-process elements, we observe two characteristic features. In the low metallicity regime ($-3 \gtrsim$ [Fe/H] $\gtrsim -1$), the models predict a steady increase of [X/Fe] from a sub-solar to solar value. At a slightly sub-solar metallicity, the models show a characteristic 'bump' feature \citep[see also][]{prantzos2018} that is qualitatively consistent with the observed data.} The predicted strength of the bump varies between s-process elements and is more prominent in heavy s-process elements, such as Ba, La, and Ce. The increasing part of the feature is associated with the transition from the dominant production of s-process elements in CCSN to AGB stars, with the latter having a slightly higher [X/Fe] ratio. The decreasing part is due to the metallicity-dependent yields of AGB stars, {which} produce less s-process elements at higher metallicity (see Figure~\ref{ls-hs-agb} and discussion in \textsection4.2) {and in part due to the production of Fe by SN Ia}. For the heaviest elements in the sample, Nd and Eu, the {the effect of varying the SFH values on their GCE tracks is very small}. {However, it is worth noting that the predicted Nd abundances tend to be lower than those observed in Gaia-ESO at [Fe/H]$\lesssim -0.2$. This mismatch might point either to an overestimated Nd abundance in the Gaia-ESO data, {incorrect yields of Nd by AGB stars,} and/or an underestimated r-component of Nd produced {in other sites}. 
In summary, the impact of the SFH on the GCE tracks of neutron-capture elements is limited. This is easy to understand, as the SFH regulates the overall enrichment speed, while the individual abundance ratios [X/Fe] - at a given metallicity - are primarily determined by the relative contributions of different sites of neutron-capture processes. 

From Fig. \ref{sfh_inflow}, it is clear that the timescale of gas accretion has an even weaker impact on the GCE tracks. In particular, this is true at sub-solar metallicities, where the models with different $\tau_{\rm inflow}$ are nearly {indistinguishable. Only around} the solar metallicity the models start to deviate from each other. Models with shorter gas accretion timescales experience less dilution, which results in a more extended ({in [Fe/H]}) 
bump feature for La, Ce, and Nd. For all other elements, including the $\alpha$-elements (O, Ca), light s-process (Y,Zr), and r-process (Eu), no significant difference is seen throughout the entire metallicity regime covered by the Galactic disc, except the most metal-rich population, where the shorter inflow timescale allows to attain significantly greater [X/Fe] ratios at a given metallicity, resulting into a pronounced upturn in the GCE tracks at [Fe/H] $\gtrsim +0.2$. However, such upturns are not observed in the data \citep[e.g.,][]{battistini2016,delgado2017,forsberg2019}, ruling out the corresponding parameter space of the models. {Our basic model that assumes $\epsilon_{*}=0.03$ and $\tau_{\rm inflow}=7$~Gyr provides an overall best match to the observed distributions and will therefore be used for the subsequent analysis.} 

{We remind the reader that in case of more complicated multi-phase SFH, which have been proposed to explain the observed complex age-chemical structure \citep[e.g.,][]{haywood2018,lian2020b}, the impact of the SFH on the enrichment of neutron-capture elements might be more complicated. Yet}, as we will show later, the observed pattern in [X/Fe]-[Fe/H] is mainly due to the nature (e.g., metallicity-dependent yields, DTD) of different sites of neutron-capture process. Therefore the main goal of this work, i.e. understanding the observational constraints from chemical abundance distribution on the sites of neutron-capture process, will not be hindered by adopting a simplified SFH. 
\subsection{s-process}
\begin{figure}
	\centering
	\includegraphics[width=8cm]{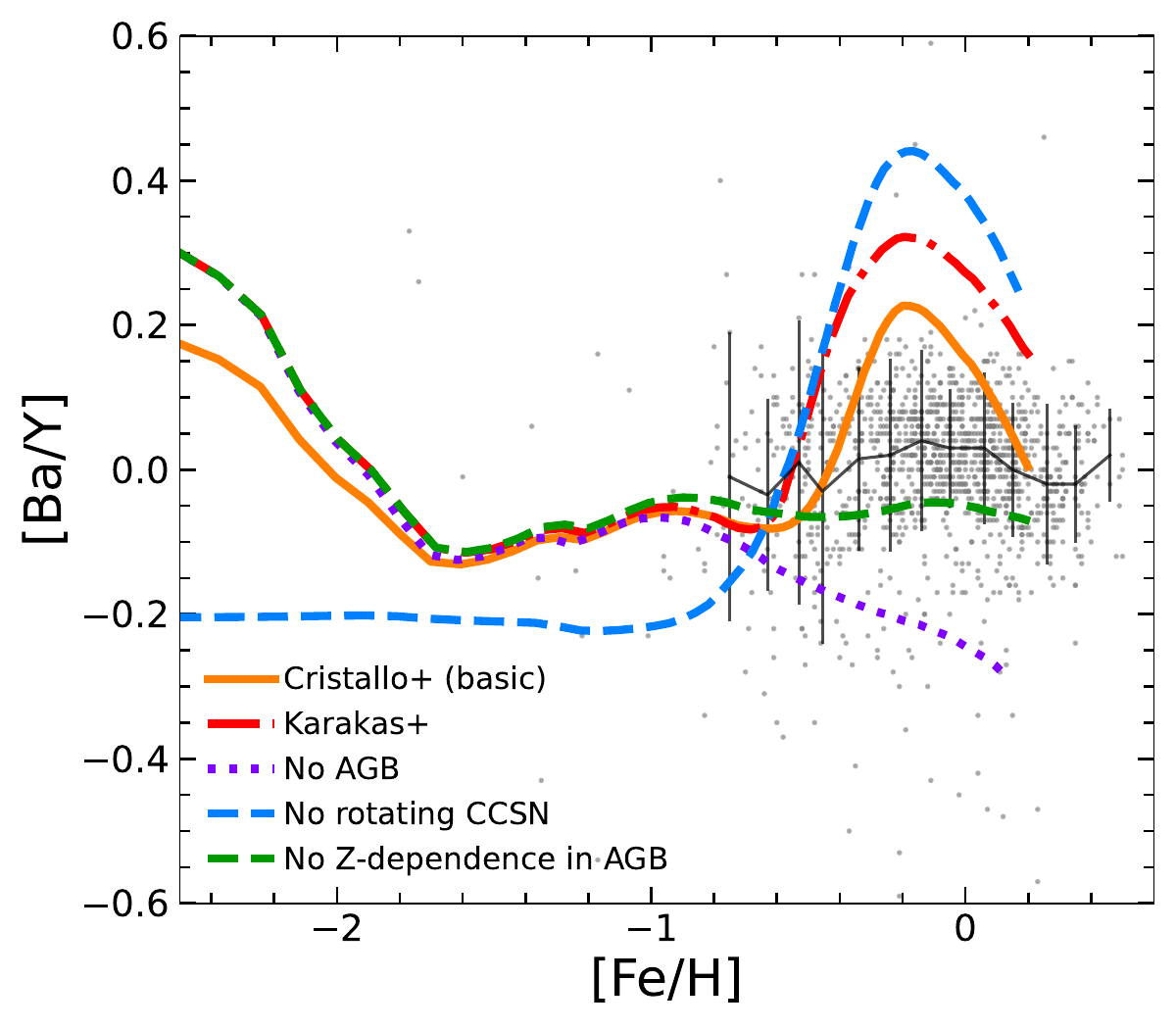}
	\caption{Relative abundance ratio between heavy {(Ba) and light (Y)} s-process elements as a function of [Fe/H]. Four GCE models with three test models deviating from the basic model in three aspects are included, one without rotating CCSN (blue dashed), one without AGB contribution (purple dotted), and one with AGB yields from Karakas et al. (magenta dash-dotted).} 
	\label{ls-hs-omega}
\end{figure} 

\begin{figure*}
	\centering
	\includegraphics[width=15cm]{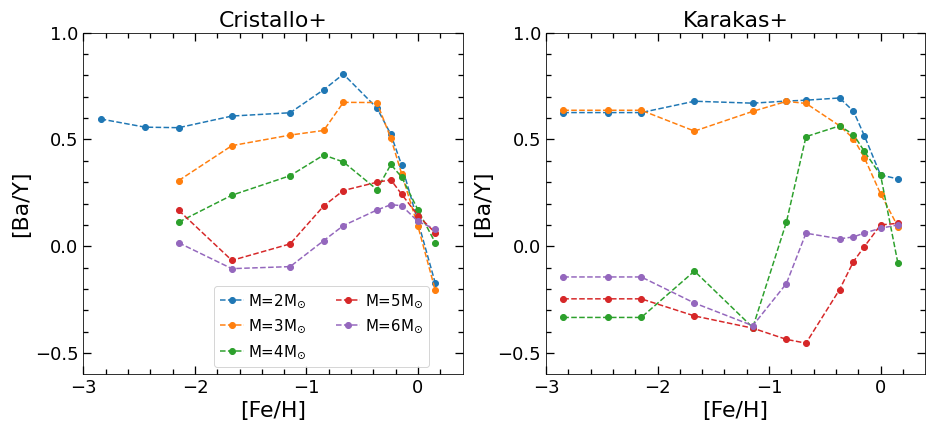}
	\caption{Heavy-to-light s-process abundance ratio ([hs/ls]) parameterized by the ratio [Ba/Y] in AGB yields from \citet{cristallo2015} (left panel) and Karakas et al. (right panel). Different masses of AGB stars are shown in different colours. {See text}.} 
	\label{ls-hs-agb}
\end{figure*} 


One of the most interesting features of s-process in the GCE plane of [s/Fe]-[Fe/H] is the weak bump at sub-solar metallicity ([Fe/H]$\sim-0.2$). This feature is clearly seen in Ba and perhaps also in Y in our Gaia-ESO sample (Fig.~\ref{xfe-feh-ges}), but not in other s-process elements, possibly because of sparse measurements at low metallicity. As discussed above, the GCE models also predict such a bump feature, which is stronger in heavy s-process elements than in the light ones. 

Figure~\ref{ls-hs-omega} shows the evolution of [Ba/Y] as a function of metallicity. {We use [Ba/Y] as a proxy for the ratio of heavy to light s-process elements, because these two elements can be more reliably measured in our stellar sample. It can be seen that this abundance ratio is close to the solar value and only a weak metallicity dependence is observed}. To explore the origin of the {[Ba/Y]} feature around the solar [Fe/H] values, following \citet{prantzos2018}, we first run four test models in addition to the basic model, one excluding the AGB contribution, one with the AGB yields adopted from Karakas et al., one with metallicity-independent AGB yields {(fixed at solar metallicity)}, and one with only non-rotating massive stars. {The resulting 4 tracks are also depicted in Fig.\ref{ls-hs-omega}.} {AGB stars make a significant contribution to the enrichment of s-process elements at [Fe/H]$\sim$-0.6 and dominate the production of s-process elements at higher metallicity}. The test model with {the} metallicity-independent AGB yields confirms that the strong metallicity dependence of [Ba/Y] originates from the metallicity dependence of s-process yields of AGB stars. 

The left-hand panel of Figure~\ref{ls-hs-agb} shows the [Ba/Y] ratio of the yields of individual AGB stars from \citet{cristallo2015} as a function of metallicity and mass. 
The steep decrease of [Ba/Y] {at high} metallicity in low-mass AGB stars is because in more metal-rich {stars} the neutron-to-seed ratio is lower.
The absolute [Ba/Y] ratio, {turnover} metallicity, and the slope of decrease beyond the peak varies between AGB stars with different masses. Intermediate mass ({4-6 ${\rm M_{\odot}}$}) AGB stars tend to reach lower [Ba/Y] ratio at a {sub-solar} metallicity and show a shallower decrease afterwards. The main {nuclear} reaction that produces free neutrons in thermally pulsating AGB stars is ${\rm ^{13}C(\alpha,n)^{16}O}$. The mass dependence of [Ba/Y] is mainly due to a lesser s-process efficiency in more massive stars that have a smaller $^{13}{\rm C}$ reservoir, an additional neutron source (${\rm ^{22}Ne(\alpha,n)}$) that favours first s-process peak elements, and smaller He-shell and larger envelopes which make s-process element enrichment more difficult \citep{cristallo2009,fishlock2014}. 
While the predicted metallicity dependence of the {corresponding GCE model tracks} is a combined result of the all AGB stars that have {occurred in a stellar population up to a given time step}, the prominent metallicity-dependence of [Ba/Y] in the models is mainly driven by low-mass AGB stars. The mismatch between the observed amplitude of the [Ba/Y] at {a slightly sub-solar metallicity, [Fe/H] $\sim -0.1$ dex} and the GCE model thus suggests that the s-process efficiency in low-mass AGB stars is {possibly overestimated \citep{magrini2021}.} 
{One possibility to lower the s-process efficiency in AGB stars is to include magnetic fields.  \citet{vescovi2020} have calibrated those models to meteoritic data, finding that they could reduce the [Ba/Y] ratio by $\sim0.2$~dex for a 2~${\rm M_{\odot}}$ AGB star at solar metallicity.} 


To further test whether our result is robust to the choice of {yields}, we adopt new AGB yields from a series of studies led by the group of A. Karakas. These yields update the data presented in \citet{karakas2010}. The predicted evolution of [Ba/Y] in Fig.~\ref{ls-hs-omega} is similar between the models with AGB yields from Karakas et al. and \citet{cristallo2015}, albeit that the model with Karakas's yields predicts slightly higher maximum [Ba/Y] ratio. The right-hand panel of Fig.~\ref{ls-hs-agb} shows the behavior of [Ba/Y] calculated using the data from Karakas et al. Although the abundance ratios show a somewhat different behavior compared to the values based on \citet{cristallo2015} data, especially for the intermediate mass AGB stars at low [Fe/H], the overall metallicity dependence is {broadly consistent \citep[see also][]{Karakas2016, karakas2018}}. In both cases, the [Ba/Y] ratios start with a mild increase from low metallicity and switch to a rapid decrease at slightly sub-solar metallicity. 
{This confirms that the observed trend of [Ba/Y] ratio at [Fe/H]$\gtrsim -0.6$ is driven by the metallicity dependence in the neutron capture efficiency in AGB stars.}

Observational evidence for the need of contribution from rotating massive stars is the observed relatively low heavy-to-light s-process abundance ratio around solar [Fe/H] \citep{prantzos2018}. This idea is supported by our work, albeit contingent on the quality of the current AGB yields. The GCE model with the yield contributions from non-rotating massive stars only shows a stronger deviation from the observed [Ba/Y] ratio (Fig.~\ref{ls-hs-omega}). {In addition, the GCE track calculated using yields from non-rotating massive stars predicts lower [Ba/Y] at [Fe/H]$<-1$. This peculiar behaviour can be verified by future observations that {would allow accurate [Ba/Y] measurements in the metal-poor regime, such as the high-resolution 4MOST surveys of the Galaxy \citep{Christlieb2019,Bensby2019}.} 
%

\subsection{r-process}

Since the only r-process element available for a large sample of Gaia-ESO stars is Eu, we focus on the [Eu/Fe]-[Fe/H] distribution to explore the observational constraints on the sites of r-process elements. 
In addition to the elemental abundance, inspired by \citet{cote2019}, we also explore the constraints on the DTD of NSMs from the estimate of coalescence time of known NS-NS binaries in the Milky Way. 
In this work we focus on two most widely discussed sites of r-process, NSM and MRSN. 

\subsubsection{NSM only}

\begin{figure*}
	\centering
	\includegraphics[width=16cm]{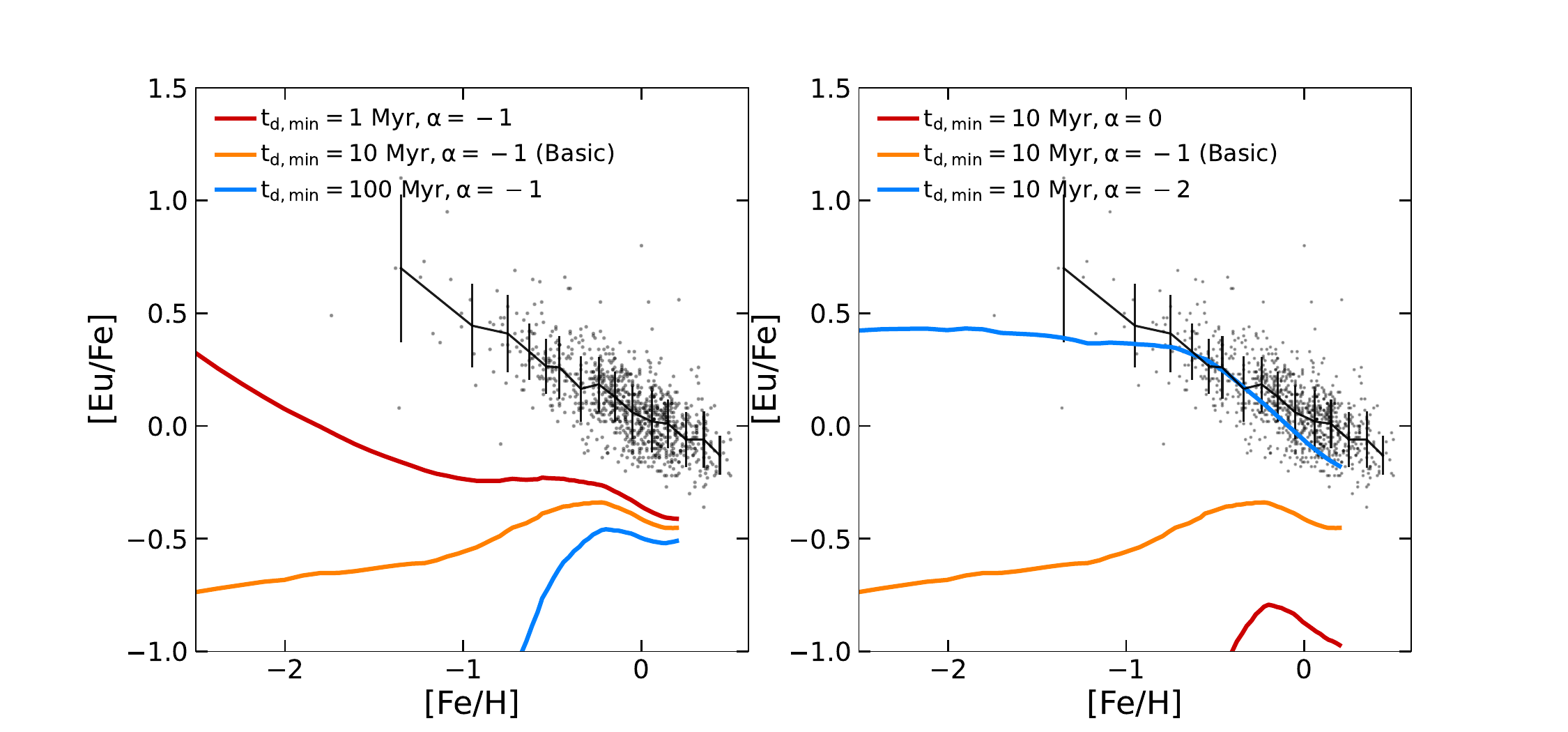}
	\caption{[Eu/Fe]-[Fe/H] evolution predicted by GCE models with NSM as the only sites of r-process. NSM DTD is characterised by the maximum delay time ($\tau_{\rm max}$) and the slope of power-law DTD ($\alpha$). Different NSM DTDs with different $\tau_{\rm max}$ (left) and $\alpha$ (right) are assumed as illustrated in the legend.} 
	\label{eufe-feh-nsm-omega}
\end{figure*} 

\begin{figure}
	\centering
	\includegraphics[width=9cm]{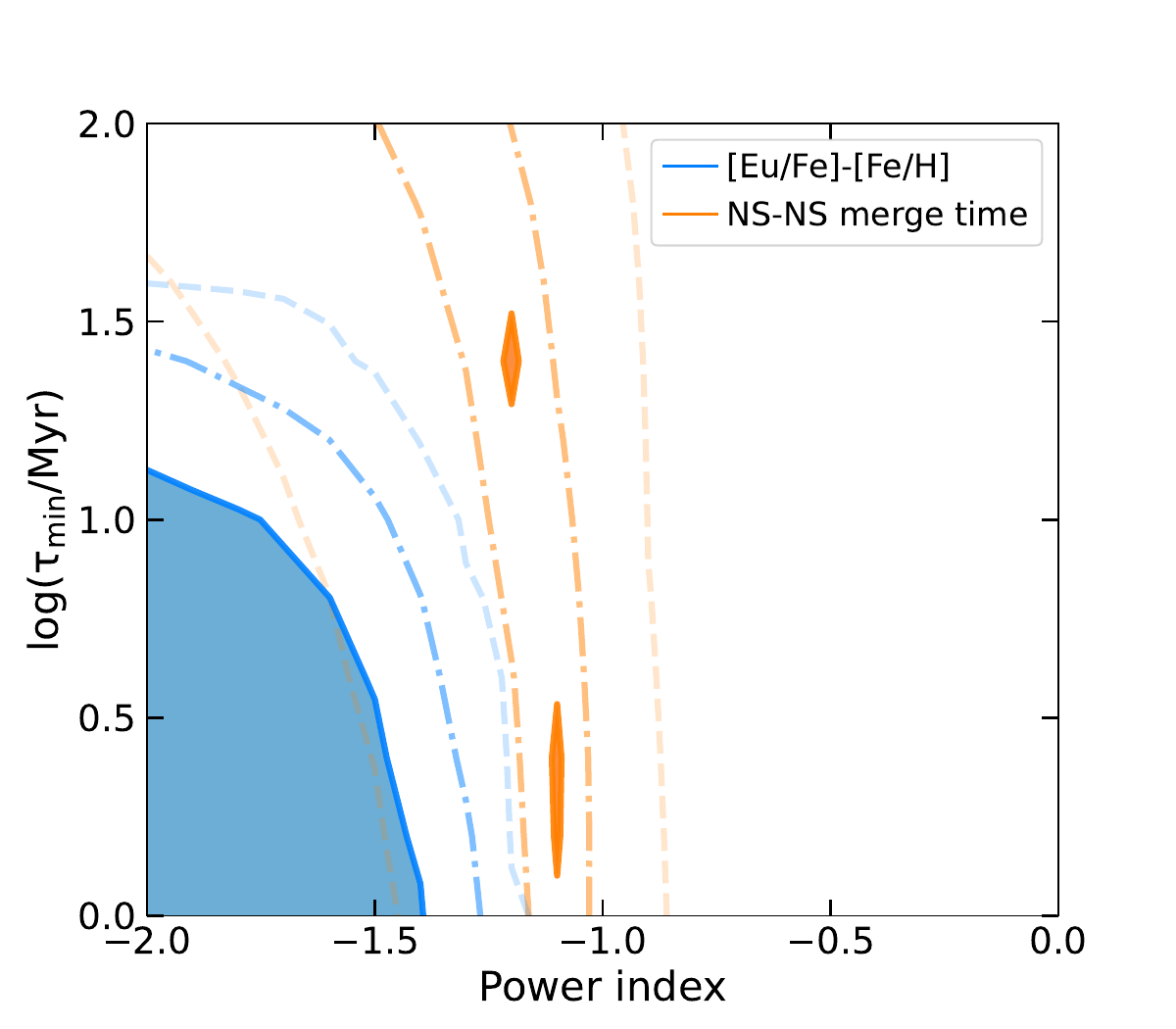}
	\caption{Probability distribution of the DTD power index and maximum delay time of NSM as constrained by [Eu/Fe]-[Fe/H] relation (blue) and coalescence time of seven confirmed NS-NS binary systems from \citet{tauris2017} (orange). The solid, dash-dotted and dashed curves denote the iso-probability contours of probability at levels of 0.9, 0.5, and 0.05. {The shaded regions indicate the parameter space with probability above 0.9.}} 
	\label{tau-slp-NSM-obs}
\end{figure}



Detection of r-process element signatures in the spectrum of kilonova \citep[e.g.,][]{chornock2017} associated with GW170817 \citep{Abbott2017a} serves as a direct evidence that NSMs are an  important site of r-process. Based on the current estimate of the rate and mass ejection of NSMs, albeit with large uncertainties, it is suggested that NSMs alone may be able to produce the amount of r-process elements observed in the Milky Way \citep{Abbott2017a}. Here we first explore the scenario that NSMs are the only source of Eu. 

The DTD of NSMs is a critical property that has a strong impact on the enrichment of r-process {elements}. Many {studies} found that SNe Ia follow a power-law DTD in a form of $t^{-1}$ (\citealt{maoz2014} and reference therein). Such a DTD is also found to be consistent with short GRB observations \citep{fong2017}, although the statistics is still low. Population synthesis models that follow the evolution of binary systems involving stars and compact remnants predict the DTD of NSMs also in the form of $t^{-1}$ \citep{cote2019}. Interestingly, a significant excess of rapidly merging NS-NS systems {in comparison to the} $t^{-1}$ DTD 
are found by \citet{beniamini2019}. 
In this  work we adopt a power-law DTD for NSMs as follows,
    \[ {\rm DTD}(\tau) \propto  \begin{cases} 
      0 & \tau<\tau_{\rm min} \\
      \tau^{-\alpha} & \tau_{min}<\tau<\tau_{\rm max} \\
      0 & \tau>\tau_{\rm max}
   \end{cases}
     \]   
and {we} set the minimum delay time ($\tau_{\rm min}$) and the power index ($\alpha$) of the DTD as free parameters. 
The maximum delay time ($\tau_{\rm max}$) is set to be 10$^6$~Gyr, according to the maximum merging time of NS-NS systems in \citet{beniamini2019}. 
Figure~\ref{eufe-feh-nsm-omega} shows the prediction of GCE models with various $\tau_{\rm min}$ (left-hand) and $\alpha$ (right-hand) in comparison with the observed [Eu/Fe]-[Fe/H] distribution of our Gaia-ESO sample. The observed [Eu/Fe]-[Fe/H] distribution shows a clear anti-correlation at [Fe/H]$>$-1, consistent with earlier observations \citep[e.g.,][]{battistini2016,delgado2017,guiglion2018,forsberg2019}. This rapid decrease of [Eu/Fe] is also seen in [$\alpha$/Fe] and suggests that the enrichment of Eu occurs, similar to $\alpha$ elements, on a much shorter {timescale than that of Fe}. The flattening at [Fe/H]$<$-1 reported in some previous works \citep[e.g.,][]{zhao2016} is not seen here, {but our sample is too scarce to draw any conclusion.} 

In the left panel of Fig.\ref{eufe-feh-nsm-omega}, when taking $\alpha= -1$, which is found in the case of SNe Ia \citep{maoz2012} and possibly short GRBs \citep{fong2017}, none of the models regardless of $\tau_{\rm min}$ match the data sufficiently well. 
All models fail to reproduce the {observed  evolution of [Eu/Fe] over the entire metallicity range probed by the Gaia-ESO sample}.  
Only at [Fe/H] $\gtrsim -0.4$, the model with a short $\tau_{\rm min}$ of 1~Myr predicts a steeper slope, {at least qualitatively} comparable to the data. In the right panel, it can be seen that the power index of the DTD has a much greater impact on the evolution of the trend in [Eu/Fe]-[Fe/H] plane. {The model with a DTD power index of $-2$ results in dramatically higher [Eu/Fe] {ratios} at [Fe/H] $\lesssim -1$ and a steep decrease of [Eu/Fe] at [Fe/H] $\gtrsim -0.8$, that matches well the observed evolutionary trend of [Eu/Fe]-[Fe/H]. The steep decrease of [Eu/Fe] predicted by the model with a steep power law of the DTD is because in this case the delay times of NSMs are significantly shorter than those of SNe Ia. The need for a short delay time of NSMs or an alternative site  to explain the abundance pattern of r-process elements has been suggested in many recent studies \citep[e.g.,][]{matteucci2014,cote2019,simonetti2019,kobayashi2022,vanderswaelmen2022}.}
  
However, such short delay time of NSMs might be in tension with other independent observations, which suggest a significant fraction of long delayed ($>$1~Gyr) NSMs \citep{cote2019}. 
Among the confirmed NS-NS binary systems in the Milky Way compiled by \citet{beniamini2019}, 
15 systems have an estimate of coalescence time and spin-down age. 
Note that 7 (or 8, if considering spin-down time) of these 15 systems have a coalescence time longer than 1~Gyr. In addition, a significant fraction ($\sim 30\%$) of short GRBs are found in early-type galaxies with stellar population older than several Gyr on average \citep{berger2014,fong2017}. The host galaxy of GW170817 is also an early-type galaxy. This NSM event likely has a delay time more than $\sim3$~Gyr \citep{pan2017,blanchard2017}. More detections of NSMs and their host galaxies in the future will provide crucial constraints on the DTD of NSMs \citep{mccarthy2020}. 

To quantify the constraints on the DTD of NSMs from {elemental} abundances and other independent observations, we run a series of models with $\tau_{\rm min}$ varying from 1 to 100~Myr at a logarithmic step of 0.2~dex and $\alpha$ varying from -2 to 0 at a step of 0.1. We then confront each model with the target observational results. For the abundance distribution, we use the  median [Eu/Fe]-[Fe/H] relation binned in [Fe/H], shown as the black error bar in Fig.~\ref{eufe-feh-nsm-omega}, to balance the uneven density of stars at different metallicities. A wide bin width of 0.4~dex is adopted at [Fe/H]$<-0.8$ because of sparse observations and a bin width of 0.1~dex used at higher metallicity. For other independent observations, given that they suggest a similar behavior of NSM DTD, 
here we only use the coalescence time of the Milky Way's NS-NS binaries \citep{tauris2017}. Considering the uncertainty in the coalescence time estimate of individual NS-NS binaries, we compare the fraction of NSMs with delay time above 1~Gyr in the NS-NS binary sample (46.7\% $\pm$ 21.4\%, assuming the Poison error) with that predicted by each NSM DTD. 
We compare each model with the observations of abundances and the fraction of long delayed NSMs and then calculate the $\chi^2$ and corresponding probability. 


Figure~\ref{tau-slp-NSM-obs} shows the probability contours in $\tau_{\rm min}$ and power index $\alpha$ given the constraints from abundances (blue) and the coalescence time of NS-NS binary (orange). The solid, dash-dotted, and dashed curves correspond to the $\tau_{\rm min}$ and $\alpha$ values that can not be rejected at a probability of 0.9, 0.5, 0.05, respectively. The observed [Eu/Fe]-[Fe/H] distribution favours DTD with {the power} index less than $\sim -1.5$ and $\tau_{\rm min}$ shorter than 10~Myr. 
{In contrast, the coalescence time estimate of NS-NS binaries favors flatter DTDs with a power index only slightly lower than $-1$, which implies more NSMs with longer delay times}. {The fraction of long delayed NSMs is dependent on both $\tau_{\rm min}$ and power index, although the dependence on the latter is much stronger. There are two solutions that can match the observed fraction of long delayed NSMs, one with $\tau_{\rm min}\sim30$~Myr and power index $\sim-1.2$ and the other with much shorter $\tau_{\rm min}$ that is less than 3~Myr and slightly higher power index.}

The observed NS-NS systems are likely biased towards young systems, because older neutron stars stop emitting in the radio and therefore are difficult to find observationally. 
{Also, a significant number of NS systems are known with merger times exceeding 50 Gyr \citep{tauris2017}, and by considering mergers only we might bias the corresponding DTD constraints to steeper DTDs.}
Thus {the fraction of long delayed NSMs is expected to be higher, suggesting a flatter NSM DTD with a higher power index. This would increase the significance of the inconsistency between the DTD of NSMs constrained by the two types of observations.} 


\subsubsection{NSM+MRSN}

\begin{figure}
	\centering
	\includegraphics[width=9cm]{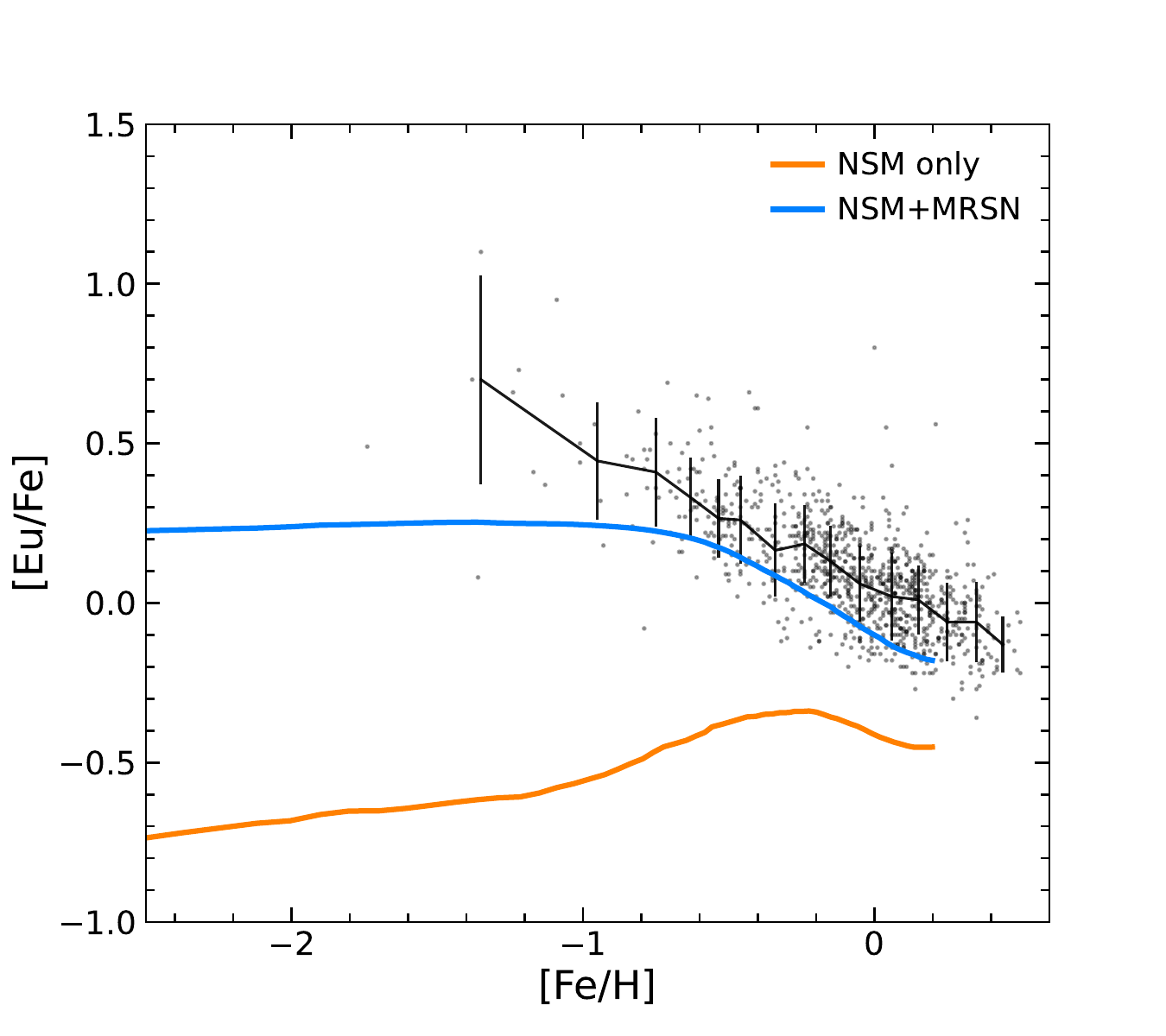}
	\caption{[Eu/Fe]-[Fe/H] evolution predicted by GCE models with both NSM and MRSN as source of r-process elements. The DTD of NSM is adopted to be the same as the basic model (i.e., $\tau_{\min}=10$~Myr and $\alpha$=-1). 
    } 
	\label{eufe-feh-nsm-mrsn-omega}
\end{figure}

\begin{figure}
	\centering
	\includegraphics[width=9cm]{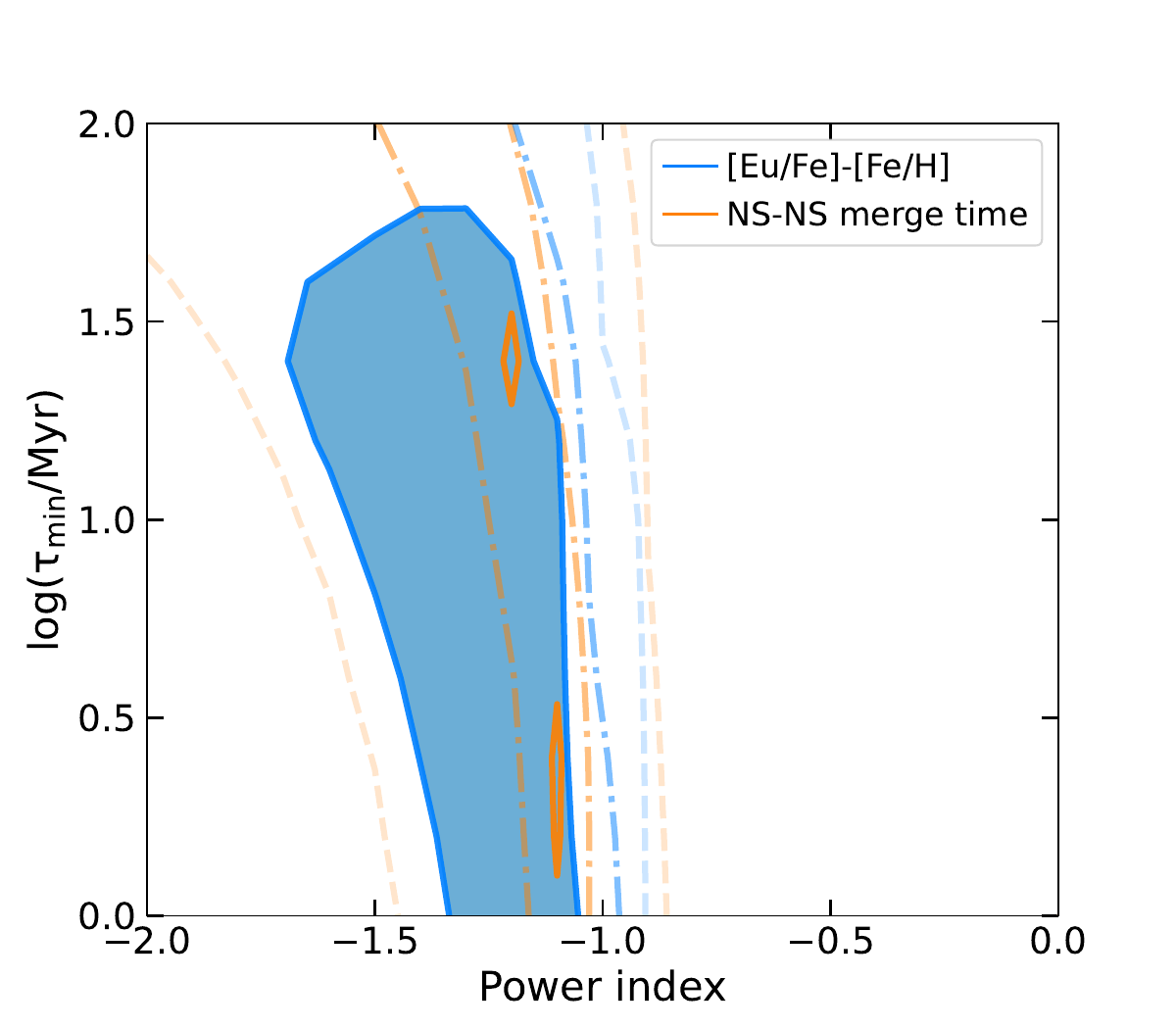}
	\caption{Same as Figure~\ref{tau-slp-NSM-obs} but replacing a very small fraction (0.25\%) of CCSN by MRSN in GCE models to meet the constraints imposed by the observed [Eu/Fe]-[Fe/H] trend.} 
	\label{tau-slp-NSM-MRSN-obs}
\end{figure}

\begin{figure*}
	\centering
	\includegraphics[width=0.9\textwidth]{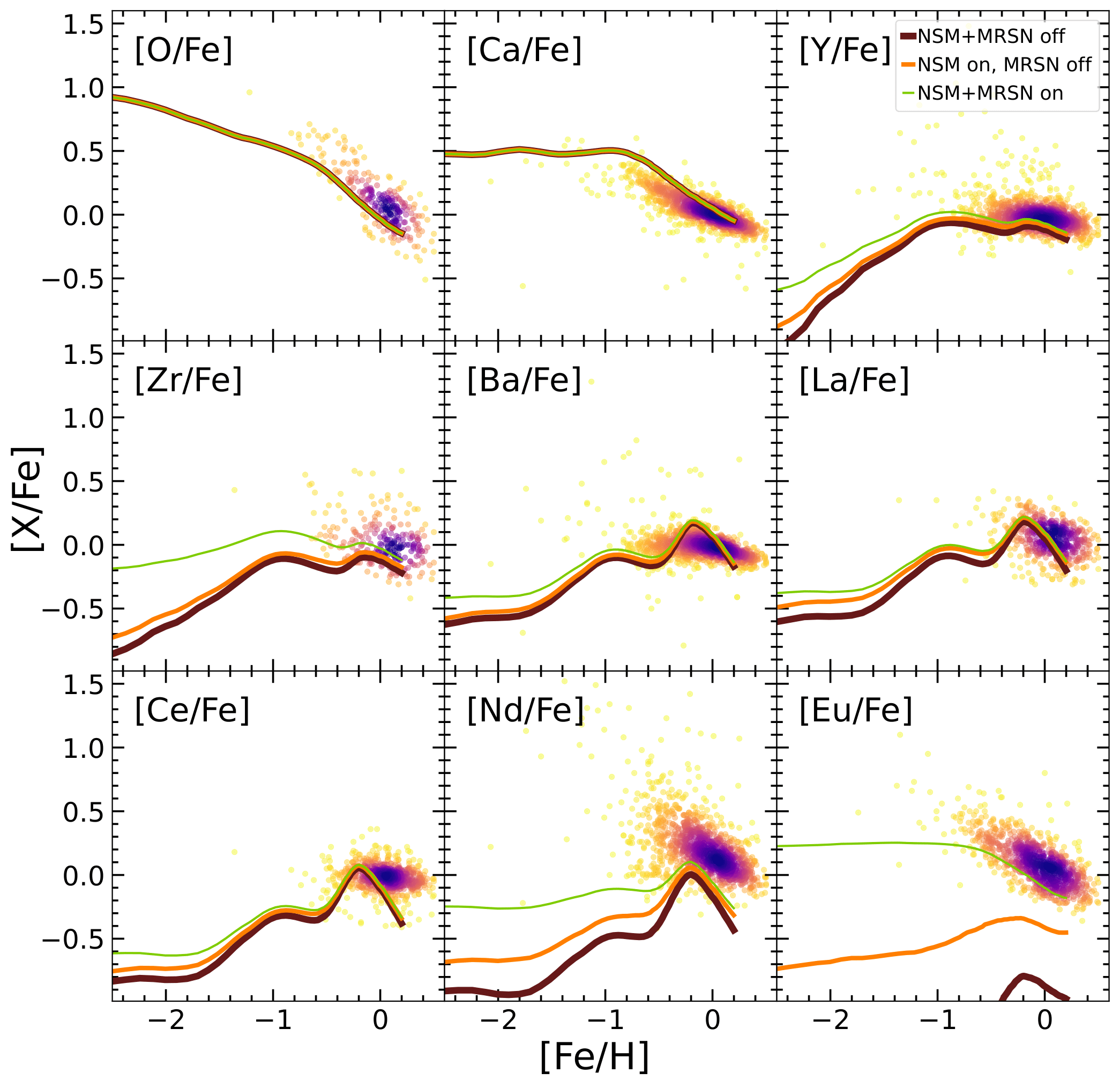}
	\caption{The impact of including NSM and MRSN, or not, on the predictions of GCE models in [X/Fe]-[Fe/H] diagram of seven neutron-capture elements in GES. Three models with different scenarios are included: no NSM or MRSN contribution (dark red), NSM only (orange), and including both NSM and MRSN (green).} 
	\label{xfe-feh-ges-sites}
\end{figure*}
To reconcile these two independent type of observations, one frequently discussed solution is to {introduce} an additional site of r-process synthesis that releases r-process elements on a much shorter timescale {compared to} NSM and {the Fe released by} SN Ia \citep{cote2019,matteucci2014}.
The most {commonly} discussed candidate of r-process site that has a short delay time is a rare type of CCSN with strong magnetic field and fast rotation, usually referred to as {magneto-rotational} supernova (MRSN, \citealt{symbalisty1984,cameron2003,nishimura2015,yong2021}). \citet{nishimura2015} calculated nucleosynthesis yields of MRSN based on a 2D magneto-hydro-dynamical simulation of a 25~${\rm M_{\odot}}$ star. Here we adopt the MRSN yields from \citet{nishimura2015} and assume that a very small fraction of massive stars that initially end up with CCSN evolve into MRSN instead. 
NSMs are also included with the same configuration as in the basic model. Note that the goal of this {experiment} is not to quantify the contribution of MRSN in the production of r-process elements, but rather to assess whether including MRSN could mitigate the tension between the model and data as discussed in the previous section or not. 

Figure~\ref{eufe-feh-nsm-mrsn-omega} illustrates the [Eu/Fe]-[Fe/H] evolution predicted by our GCE models  including MRSN's contribution. As discussed in Sect. \ref{sec:yields}, we adopt a fraction of 0.25\%~for MRSN (blue line). 
It can be seen that in comparison to the model with NSM as the only r-process site the Eu abundance is significantly enhanced by including MRSN, especially in the low metallicity regime, matching well the observed [Eu/Fe]-[Fe/H] trend over the metallicity range probed by our sample. 

To verify this result, we perform the same $\chi^2$ statistic analysis as the NSM-only models for the models considering both NSM and MRSN.  Figure~\ref{tau-slp-NSM-MRSN-obs} shows the iso-probability contours for the NSM+MRSN GCE model. 
It can be seen that {models that include MRSN can meet both the observational constraints from stellar abundances and coalescence times of NS-NS binaries within {a 1~$\sigma$} confidence level simultaneously.} With such a model, a significant fraction of Eu in the Milky Way today is synthesized in MRSN and NSMs. For example, assuming a NSM DTD with $\tau_{\rm min}=10$~Myr and $\alpha= -1$ and a fraction of $0.0025$ for MRSN, at the present day, 46.5\% of Eu is produced by MRSN, 37.3\% by NSMs, and a minor fraction of 16.1\% via s-process nucleosynthesis in AGB stars.  

{Whereas magneto-rotational supernovae clearly have a critical contribution to the production of r-process elements, such as Eu discussed above, this site is also important for other neutron-capture elements. Figure~\ref{xfe-feh-ges-sites} shows the GCE} tracks of three models with or without contributions from NSMs and MRSN. For the models with NSM contribution, we adopt a NSM DTD with $\tau_{\rm min}=10$~Myr and $\alpha= -1$. {It can be seen that, similar to Eu but to a lesser extent, inclusion of MRSN significantly enhances the abundances of lighter n-capture elements at low metallicity, especially Zr and Nd. This has an impact on the r-process fraction of these elements, which are dominated by s-process at present day}. The impact at the high metallicity end is negligible, suggesting that, in this example model, AGB stars are still the major source of s-process elements.
%
%



\section{Summary and conclusions}
In this work we investigate the observational constraints of chemical abundances {of Galactic stars on the sites of neutron-capture nucleosynthesis. We use the new data from the Gaia-ESO large spectroscopic survey, specifically O, Ca, Fe, Y, Zr, Ba, La, Ce, Nd, and Eu and the OMEGA+ chemical evolution model of the Galaxy. In addition to the yields, we explore the influence of basic parameters in the standard 1-zone GCE models, including the star formation efficiency and the inflow timescales, and explore their impact on the predicted Galactic chemical enrichment tracks.} 

{For the s-process}, we use the abundance ratio between Ba and Y 
to trace the neutron-capture efficiency in the s-process sites. This abundance ratio shows a weak metallicity dependence that is likely caused by the metallicity dependence in AGB yields. 
While our model provides a good match to the overall trend of the evolution of [Ba/Y], the predicted {metallicity dependence is clearly too strong} compared to the observations. By inspecting the yields of individual AGB stars, we find that this mismatch is caused by the strong metallicity-dependent yields of [Ba/Y] in low-mass AGB stars. To better match the observed [Ba/Y], we anticipate that a lower s-process efficiency is needed in low-mass AGB stars. {One possibility to lower the s-process efficiency is to include magnetic fields in AGB stars.}

For the r-process, we explore the observational constraints from the element abundances (Eu) and the coalescence time of known NS-NS binary systems, although the latter might be subject to uncertainties that are currently difficult to account for.  
When assuming NSMs as the only site of r-process, we find these two independent observational {constraints favour} different delay time distributions for NSMs. While a steep DTD with {the power index lower than $\sim -1.5$ is required to match the steep decline of [Eu/Fe]} at [Fe/H] $\gtrsim -1$ dex, a shallower DTD with a power index $\sim -1$ is needed to explain the significant fraction of long coalescence time ($>1$~Gyr) of NS-NS binaries and short GRBs in early-type galaxies. 

{Our results suggest that an additional site of r-process with a short delay time is needed. We also explore the possibility of MRSN and find that including MRSN helps to alleviate the need for uncomfortably short delay time of NSMs. Our model including MRSN and NSMs with the standard DTD of NSM in the form of $t^{-1}$ is able to simultaneously explain the observed rapid decrease of [Eu/Fe] in Galactic stars and the notable frequency of long delayed NSM as indicated by the coalescence time of NS-NS binaries and host galaxy properties of GW170817 and short GRBs. It is worth pointing out that in this model a significant fraction of Eu at present day comes from NSMs ($46.5\%$), while a large fraction ($37.3\%$) is produced by MRSN, which however, dominated the Eu production at [Fe/H]$\lesssim -1$.}



{The analysis of neutron-capture element enrichment} in the high-metallicity regime relies on the knowledge of the early r-process enrichment in short-delay sites, e.g., CCSN and MRSN. To deepen our understanding of the origins and evolution of neutron-capture elements, a critical step in the future is to extend the chemical abundance analysis to a lower metallicity regime {and to improve the accuracy of element abundances}. We expect that the next generation  massive stellar spectroscopic surveys, such as 4MOST and WEAVE, will provide homogeneous measurements of neutron-capture element abundances for large samples of stars at [Fe/H] $\lesssim -1$ dex, which will serve as the ideal datasets to study the early enrichment of neutron-capture elements.  

\section*{Acknowledgements}
Based on observations made with ESO Telescopes at the La Silla or Paranal Observatories under programme ID(s) 072.D-0019(B), 072.D-0309(A), 072.D-0337(A), 072.D-0406(A), 072.D-0507(A), 072.D-0742(A), 072.D-0777(A), 073.C-0251(B), 073.C-0251(C), 073.C-0251(D), 073.C-0251(E), 073.C-0251(F), 073.D-0100(A), 073.D-0211(A), 073.D-0550(A), 073.D-0695(A), 073.D-0760(A), 074.D-0571(A), 075.C-0245(A), 075.C-0245(C), 075.C-0245(D), 075.C-0245(E), 075.C-0245(F), 075.C-0256(A), 075.D-0492(A), 076.B-0263(A), 076.D-0220(A), 077.C-0655(A), 077.D-0246(A), 077.D-0484(A), 078.D-0825(A), 078.D-0825(B), 078.D-0825(C), 079.B-0721(A), 079.D-0178(A), 079.D-0645(A), 079.D-0674(A), 079.D-0674(B), 079.D-0674(C), 079.D-0825(B), 079.D-0825(C), 079.D-0825(D), 080.B-0489(A), 080.B-0784(A), 080.C-0718(A), 081.D-0253(A), 081.D-0287(A), 082.D-0726(A), 083.B-0083(A), 083.D-0208(A), 083.D-0671(A), 083.D-0682(A), 083.D-0798(B), 084.D-0470(A), 084.D-0693(A), 084.D-0933(A), 085.D-0205(A), 086.D-0141(A), 087.D-0203(B), 087.D-0230(A), 087.D-0276(A), 088.B-0403(A), 088.B-0492(A), 088.C-0239(A), 088.D-0026(A), 088.D-0026(B), 088.D-0026(C), 088.D-0026(D), 088.D-0045(A), 089.D-0038(A), 089.D-0298(A), 089.D-0579(A), 090.D-0487(A), 091.D-0427(A), 092.D-0171(C), 092.D-0477(A), 093.D-0286(A), 093.D-0818(A), 094.D-0363(A), 094.D-0455(A), 171.D-0237(A), 187.B-0909(A), 188.B-3002(A), 188.B-3002(B), 188.B-3002(C), 188.B-3

AS acknowledges grants PID2019-108709GB-I00 from Ministry of Science and Innovation (MICINN, Spain), Spanish program Unidad de Excelencia Mar\'{i}a de Maeztu CEX2020-001058-M, 2021-SGR-1526 (Generalitat de Catalunya), and support from ChETEC-INFRA (EU project no. 101008324).

MB and JL are supported through the Lise Meitner grant from the Max Planck Society. MB acknowledges support by the Collaborative Research centre SFB 881 (projects A5, A10), Heidelberg University, of the Deutsche Forschungsgemeinschaft (DFG, German Research Foundation). This project has received funding from the European Research Council (ERC) under the European Union’s Horizon 2020 research and innovation programme (Grant agreement No. 949173).

\section*{Data Availability}

The data underlying this article is from the public data release of the Gaia-ESO survey.


\bibliographystyle{mnras}
\bibliography{Jianhui}{}

\appendix

\section{GCE model prediction for all heavy elements}

\begin{figure*}
	\centering
	\includegraphics[width=0.99\textwidth]{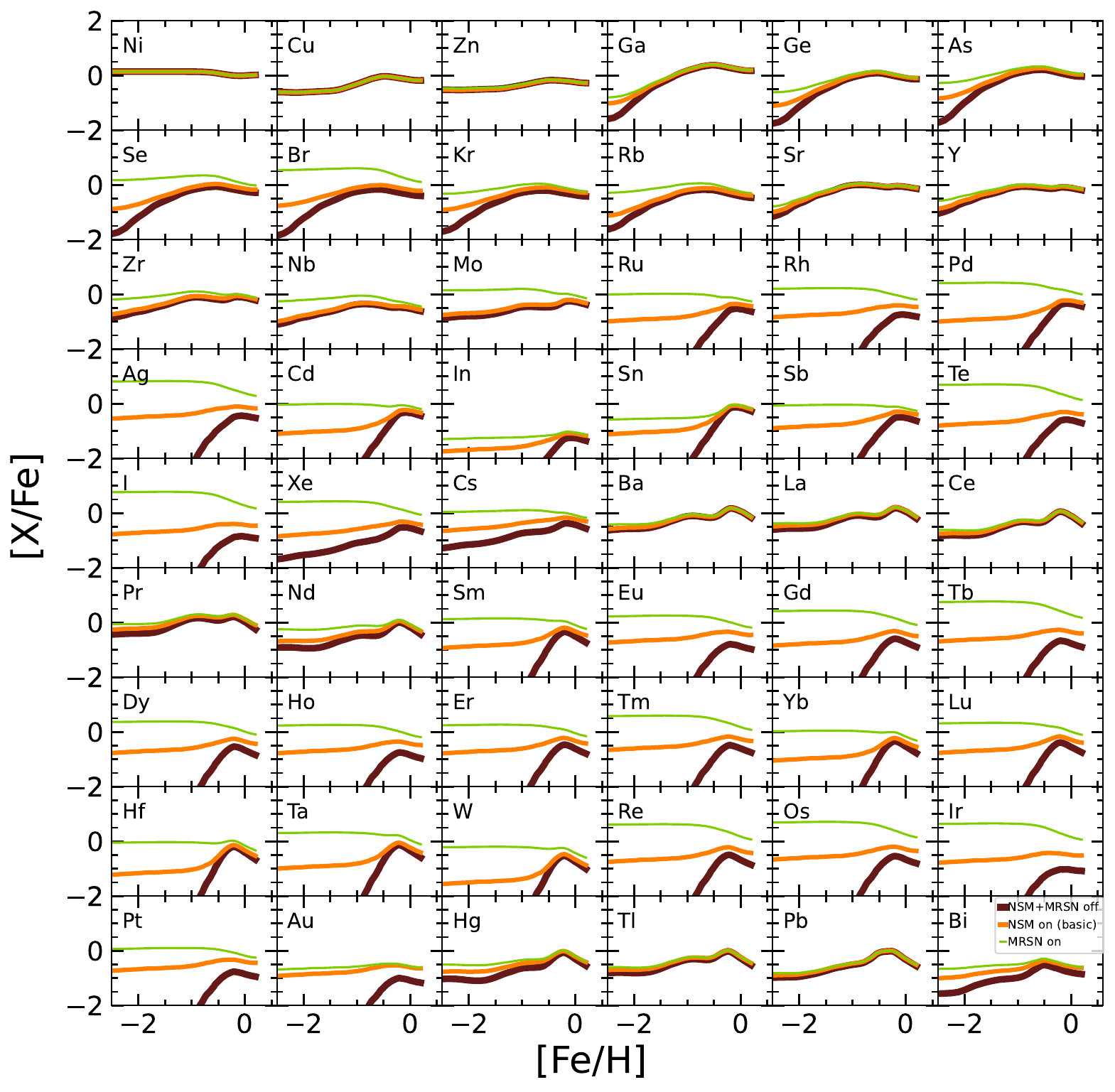}
	\caption{The impact of including NSM and MRSN, or not, on the predictions of GCE models in [X/Fe]-[Fe/H] diagram for most elements after Fe. Colouring is the same as in Fig. \ref{xfe-feh-ges-sites} for three models with different scenarios: no NSM or MRSN contribution (dark red), NSM only (orange), and including both NSM and MRSN (green).} 
	\label{xfe-feh-appendix}
\end{figure*}

. 

\end{document}